\documentclass[12pt]{article}
\usepackage{graphicx}  % standard LaTeX graphics tool;
\usepackage{amsmath}   % For getting proper math eqns
\usepackage[compress]{cite}
\usepackage{amssymb}   % Used for getting various symbols eg. gtrsim, lesssim etc.
\usepackage{bm} % For bold math (esp of lower greek letters)
\usepackage{dcolumn}% Align table columns on decimal point
\usepackage{color}
\usepackage{mathrsfs}
\usepackage{ulem}
\usepackage{amsfonts}
\usepackage[utf8]{inputenc}
\usepackage{varioref}
\usepackage{csquotes}
\usepackage[caption=false]{subfig}
\RequirePackage[colorlinks,citecolor=blue,urlcolor=magenta,linkcolor=blue]{hyperref}
\def \MPD {Mathisson-Papapetrou-Dixon}
\def \SSC {spin supplementary condition}
\def \TD {Tulczyjew-Dixon}
\def\P{Mathisson-Pirani}

\def \S {Schwarzschild}

\def \S {Schwarzschild}
\def \OKS {Ohashi-Kyrian-Semer\'{a}k}

\labelformat{section}{Section-(#1)} 
\labelformat{subsection}{Section-(#1)} 
\labelformat{subsubsection}{Section-(#1)}
\labelformat{subsubsubsection}{Section-(#1)}
\labelformat{figure}{Fig.~(#1)} 
\labelformat{equation}{Eq.~(#1)}
\labelformat{subfigure}{Fig.~(\thefigure#1)} 
\labelformat{table}{Tab.~#1} 
\labelformat{appendix}{Appendix #1}

\title{Extended bodies moving on geodesic trajectories}
\author{Sajal Mukherjee \footnote{mukherjee@asu.cas.cz}~$^{1,2}$, Georgios Lukes-Gerakopoulos\footnote{gglukes@gmail.com}~$^{2}$ \\
and Rajesh Kumble Nayak \footnote{rajesh@iiserkol.ac.in}~$^{1,3}$,\\
$^{1}${\small{Department of Physical Sciences, IISER-Kolkata, Mohanpur-741246, India}}\\
$^{2}${\small Astronomical Institue of the Czech Academy of Sciences,}\\
{\small Bocni II 1401/1a, CZ-14100 Prague, Czech Republic}\\
$^{3}$ {\small Center of Excellence in Space Sciences India, IISER-Kolkata, India}\\
}

\begin{document}
\maketitle
\begin{abstract}
This work investigates whether an extended test body obeying the \MPD~equations under the \OKS~\SSC~can follow geodesic trajectories in curved spacetimes. In particular, we explore what are the requirements under which pole-dipole and pole-dipole-quadrupole approximated bodies moving in the \S~or Kerr spacetimes can follow equatorial geodesic trajectories. We do this exploration thoroughly in the pole-dipole case, while we focus just on particular trajectories in the pole-dipole-quadrupole case. Using the \OKS~\SSC~to fix the center of the mass of a pole-dipole body has the advantage that the hidden momentum is eliminated. This allows the four-velocity to be parallel to the four-momentum, which provides a convenient framework for our investigation. We discuss how this feature can be recovered at a pole-dipole-quadrupole approximation and what are the consequences.
\end{abstract}

\section{Introduction}

The motion of an extended test body has long and extensive history starting with the seminal works of Mathisson \cite{Mathisson:1937zz}, Papapetrou \cite{Papapetrou:1951pa,corinaldesi1951spinning} and Dixon \cite{dixon1964covariant,dixon1970dynamics,dixon1970dynamics2,dixon1974dynamics} up to more recent treatments \cite{Semerak:1999qc,Kyrian:2007zz,Saijo:1998mn,Hartl:2003da,Barausse:2009xi,Taracchini:2013rva,Jefremov:2015gza,Mukherjee:2018zug,Mukherjee:2018bsn,Mukherjee:2018kju,Semerak:2015dza,Deriglazov:2018vwa,Costa:2014nta,Costa:2017kdr,Lukes-Gerakopoulos:2014dma,PhysRevD.81.104012}. In the framework of the \MPD~equations, the extended body is often viewed as a point particle endowed with multipole  moments, since practically the whole body is represented by its center of mass called centroid with a set of multipole moments defined around it. Moreover, in the test particle limit, we ignore any back-reaction on the background and we assume that the geometry remains unaltered even if the particle is endowed with a finite size. In fact, the only corrections that one anticipates in the \MPD~framework originate from the non-vanishing coupling between the curvature and the higher order multipole moments than the mass (monopole), i.e., dipole, quadrupole and higher multipoles of the body. Due to this coupling, the particle, in general, experiences an acceleration making it deviate from a geodesic trajectory. 

There is a useful analogy between the electromagnetic field and the gravitational field helping us to understand the orbital dynamics of a test body. Even though the corresponding theories are intrinsically different, numerous useful analogies between these fields have been found over the years \cite{harris1991analogy,ciufolini1995gravitation,Ruggiero:2002hz}. In the case of an extended body, this analogy was first explored by Wald, who proved that the force on a magnetic dipole in a electromagnetic field is analogous to the force experienced by a spinning particle in a weak gravitational field \cite{Wald:1972sz}. Later the analogy between the spin precession of a gyroscope in a gravitational field and the precession of a magnetic dipole was discussed in \cite{Mashhoon:1984fj,Jantzen:1992rg,FilipeCosta:2006fz}. In recent years, this analogy have been further explored in \cite{Natario:2007pu,Costa:2012cw} and the connection between the motion of a spinning particle with its electromagnetic counterpart has been emphasized \cite{Costa:2012cy}. 

Our work focuses on a counterexample to the aforementioned analogy. Namely, contrary to the electromagnetic field case, where a moving magnetic dipole can never experience a vanishing force, in a gravitational field, a spinning particle may follow a geodesic trajectory \cite{Costa:2012cy}. Such trajectories may appear as special cases, but their existence suggests an interesting twist in the analogy. Hence, in the present work, we aim to investigate under which conditions such trajectories can be obtained. Such purely academic investigations have been already undertaken in the \MPD~pole-dipole approximation \cite{Hojman:2016mox,white2000radial,Costa:2012cy}. In particular, in Refs.~\cite{white2000radial,Hojman:2016mox} these trajectories have been studied in the \S~spacetime and it has been argued that in the case of a radially in-falling spinning particle, the spin curvature coupling identically vanishes. In Ref.~\cite{Costa:2012cy}, the existence of geodesic trajectories for a spinning particle has been considered in Kerr and Kerr-dS geometry showing that the circular geodesic in Kerr-dS can have a zero curvature coupling while the same is not true for Kerr. Hence, the existence of this special type of trajectories appears to depend on the background geometry. 

In our analysis, we explore the existence of geodesic trajectories followed by a body with dipole and quadrupole moments in the \S~and Kerr spacetimes. Our study is restricted on the equatorial plane; in the case of \S~background this is imposed by the background's geometry, while in the case of a Kerr background this is imposed by the complexity of treating generic off-equatorial orbits. In both backgrounds, we investigate two types of bodies: the pole-dipole body, in which we ignore all the higher moments than the dipole one, and the pole-dipole-quadrupole one, in which we ignore all the higher moments than the quadrupole one. By searching for the existence of geodesic orbits followed by extended bodies, we are not only able find the existence of such orbits in the pole-dipole approximation, but also exclude all the other possibilities. In the pole-dipole-quadrupole approximation, we investigate only the geodesic trajectories which we found to be compatible with the pole-dipole body. For these geodesic trajectories we couldn't find a viable solution for a pole-dipole-(spin induced) quadrupole body.

When dealing with \MPD~equations, a discussion about a \SSC~is unavoidable \cite{corinaldesi1951spinning,Pirani:1956tn,tulczyjew1959motion,Newton:1949cq}. Different \SSC s can lead to different worldlines of the same physical body \cite{Semerak:1999qc,Costa:2014nta,Costa:2017kdr}. This is intimately related with the fact that in relativity, the location of the  center of mass of an extended spinning body is frame-dependent and not defined uniquely \cite{Costa:2014nta,Steinhoff:2015ksa}. In the present work, we use the \OKS~\SSC~\cite{Ohashi03,Kyrian:2007zz,Semerak:2015dza} because in the pole-dipole approximation under this \SSC~the four-momentum and the four-velocity of the body are parallel, which is a characteristic feature of geodesic orbits not necessarily holding for the \MPD~equations. Actually, this feature does not have to hold for \OKS~when higher multipoles than the dipole are included into the extended bodies' approximation. However, the authors in \cite{Costa:2014nta} have suggested that enforcing an extra condition involving torque can allow this feature to be recovered in the pole-dipole-quadrupole approximation framework. Following this suggestion we elaborate on this condition and discuss its consequences.

The manuscript is organized as follows: \ref{sec:general} introduces the equations of motion of a pole-dipole-quadrupole particle, provides a detailed discussion about the \OKS~\SSC~and briefs the framework of the analysis that follows. The specific calculations regarding the extended bodies on geodesic orbits are carried out in \ref{sec:SchwBH} for \S~\\ and for Kerr in \ref{sec:KerrBH}. The article concludes in \ref{sec:remarks}.

\textit{Notation and Conventions:} In the paper, we have used the $(-,+,+,+)$ metric signature and set the speed of light, $c$, and gravitational constant $G$, as $c=1=G$. The \enquote*{[~]} describes the standard antisymmetric expansion given by $A^{[ \alpha }B^{\beta ]}=\dfrac{1}{2}\bigl(A^{\alpha}B^{\beta}-A^{\beta}B^{\alpha}\bigr)$. A covariant derivative is denoted as $\nabla_\mu$. Note that a partial derivative with respect to variable $z$ as $\partial_z$. A projection of a tensor on a tetrad field frame $e^{(\mu)}_\nu$, where $(\mu)$ is the tetrad index and $\nu$ is the tensor index, is denoted by putting a parenthesis around the index of the tensor, e.g. $X^{(\mu)}=e^{(\mu)}_\nu X^\nu$.

\section{The pole-dipole-quadrupole approximation and the \OKS~\SSC} \label{sec:general}

The \MPD~equations of a body in the pole-dipole-quadrupole approximation \cite{Steinhoff:2009tk,Steinhoff:2012rw}, when only gravitational interactions are considered, read 
\begin{eqnarray}
\dot{P}^{\mu}& = & -\frac{1}{2}R^{\mu}_{~\nu \alpha \beta}\mathcal{U}^{\nu} S^{\alpha \beta}-\dfrac{1}{6}J^{\alpha \beta \gamma \delta} \nabla^{\mu}R_{\alpha \beta \gamma \delta}, \label{eq:MPD_P}  \\
\dot{S}^{\mu \nu} & =& 2 P^{[\mu} \mathcal{U}^{\nu]}+\dfrac{4}{3} J^{\alpha \beta \gamma [\mu}R^{\nu]}_{~\gamma \alpha \beta },\label{eq:MPD_S}
\end{eqnarray}
where $R^{\mu}_{~\nu \alpha \beta}$ is the Riemann tensor, $S^{\alpha \beta}$ is the spin tensor, $J^{\alpha \beta \gamma \delta}$ is the quadrupole tensor, $P^{\mu}$ is the four-momentum  and $\mathcal{U}^{\mu}$ is the four-velocity, while the \enquote*{dot} defines a covariant derivative $\mathcal{U}^\mu \nabla_\mu$ with respect to the proper time. Note that \MPD~evolution equations describe only the evolution of the momentum and the spin, while the quadrupole moment is determined by the matter structure of the body.

Let us now briefly discuss the multipoles of the body in the pole-dipole-quadrupole approximatio. The mass of the body corresponding to the monopole can be defined as $m=-\mathcal{U}^\mu P_\mu$ or as $\mu^2=-P_\mu P^\mu$. These two masses are in general not equal, but it is interesting to notice that contractions between the vectors $(P^\mu-m \mathcal{U}^\mu),~\mathcal{U}^\mu,~P^\mu$ lead to the relations
\begin{align}
&(P^\mu-m \mathcal{U}^\mu)(P_\mu-m \mathcal{U}_\mu)=-\mu^2+m^2, \nonumber \\
&(P^\mu-m \mathcal{U}^\mu)P_\mu=-\mu^2+m^2,\: (P^\mu-m  \mathcal{U}^\mu)\mathcal{U}_\mu=0, \label{eq:UPcontr}
\end{align}
which imply that if $\mu^2=m^2$, then $P^\mu=m~\mathcal{U}^\mu$, since $(P^\mu-m \mathcal{U}^\mu)$ cannot be a null-vector \cite{Kyrian:2007zz}. The spin tensor is associated with the dipole moment of the body, and provides a nonzero contribution to the acceleration due to the coupling with the background Riemann tensor. An identical phenomenon occurs when the particle is endowed with the quadrupole moment. In general, the quadrupole tensor $J^{\alpha \beta \gamma \delta}$ has a more complicated and nontrivial physical description than the spin tensor \cite{Ehlers1977,Bini:2008zzf,Bini:2013uwa,Bini:2014xyr}. The total quadrupole moment is composed from different parts, i.e. the  mass quadrupole $Q^{\alpha \beta}$, the flow quadrupole $\Pi^{\alpha \beta \gamma}$ and the stress quadrupole $\tau^{\alpha \beta \gamma \delta}$ \cite{Ehlers1977}. Following the conventions introduced in  \cite{Ehlers1977},  the quadrupole components can be written as
\begin{align}
\tau^{\alpha \beta \gamma \delta}&=J^{\alpha \beta \gamma \sigma}(h^\mathcal{V})^{\delta}_{~\sigma}, \quad \Pi^{\alpha \beta \gamma} =  -J^{\alpha \sigma \beta \delta} (h^\mathcal{V})^{\gamma}_{~\delta} \mathcal{V}_{\sigma}, \quad Q^{\alpha \beta} = \dfrac{4}{3} J^{\alpha \gamma \beta \delta} \mathcal{V}_{\gamma} \mathcal{V}_{\delta},
\end{align}
for a {\it generic} time-like vector $\mathcal{V}^\mu$ satisfying $\mathcal{V}^\mu \mathcal{V}_\mu=-1$ \cite{Bini:2013uwa}, where  $(h^\mathcal{V})^{\delta}_{~\sigma}$ is a projection operator defined as 
\begin{equation}
(h^\mathcal{V})^{\delta}_{~\sigma}=\delta^{\delta}_{~\sigma}+\mathcal{V}^{\delta}\mathcal{V}_{\sigma}.
\end{equation}
This operator projects any vector on a local rest frame orthogonal to the timelike vector field $\mathcal{V}^{\alpha}$. The quadrupole components are spatial with respect to the reference vector $\mathcal{V}^\mu$ \cite{Bini:2013uwa}, i.e.
\begin{align}
\mathcal{V}_{\alpha} \tau^{\alpha \beta \gamma \delta}=\mathcal{V}_{\alpha} \Pi^{\alpha \beta \gamma}=\mathcal{V}_{\alpha} Q^{\alpha \beta}=0 .  
\end{align}
In this decomposition framework the quadrupole moment reads \cite{Harte20} 
\begin{equation}\label{eq:QuadTens}
J^{\alpha \beta \gamma \delta}=\tau^{\alpha \beta \gamma \delta}-3 \mathcal{V}^{[\alpha}Q^{\beta][\gamma}\mathcal{V}^{\delta]}-\mathcal{V}^{[\alpha}\Pi^{\beta] \gamma \delta}-\mathcal{V}^{[\gamma}\Pi^{\delta] \alpha \beta}.
\end{equation}
The presence of a dipole and quadrupole moment has an impact on the relation between the four-momentum $P^{\mu}$ and the four-velocity $\mathcal{U}^{\mu}$. Namely, these four-vectors are not, in general, parallel as they are for the geodesic motion. In fact, $P^\mu$ can be split in a part parallel $P_{\|}^\mu$ and a part orthogonal $P^{\mu}_{\rm hid}$ to $\mathcal{U}^\mu$ \cite{Costa:2014nta}, i.e.
\begin{equation}\label{eq:MomDec}
    P^\mu=P_{\|}^\mu+P^{\mu}_{\rm hid},
\end{equation}
where the orthogonal part is called \textit{hidden momentum} and is defined as
\begin{equation}
P^{\mu}_{\rm hid} =(h^\mathcal{U})^\mu_{~\nu} P^\nu ,
 \label{eq:P_Hidden}
\end{equation}
which reduces the parallel part to
\begin{equation}\label{eq:MomPar}
    P_{\|}^\mu=m\, \mathcal{U}^\mu .
\end{equation}

The dependence of the hidden momentum on the multipole moments can be seen, by contracting \ref{eq:MPD_S} with $\mathcal{U}_\nu$, i.e.
\begin{align}
 P^{\mu}_{\rm hid}=\left(K^{\mu\nu}-\dot{S}^{\mu\nu}\right)\mathcal{U}_\nu , 
\end{align}
where the second-rank anti-symmetric tensor $K^{\mu \nu}=\dfrac{4}{3} J^{\alpha \beta \gamma [\mu}R^{\nu]}_{~\gamma \alpha \beta }$ has been defined. There is, however, a way to eliminate the quadrupole hidden momentum in some cases as we discuss in the following section. 

\subsection{The \OKS~\SSC}\label{sec:OKS}

The \MPD~equations are not sufficient to close the system, a \SSC~is needed to fix the centre of mass, i.e. the centroid. In general a \SSC~ can be written as
\begin{align}
 \mathcal{V}_\mu S^{\mu\nu}=0.
\end{align}
As was already mentioned, in our analysis the \OKS~\SSC~shall be employed. According to this~\SSC \cite{Kyrian:2007zz}, a vector $\mathcal{V}^\mu=w^\mu$, i.e. $S^{\mu\nu}w_\mu=0$, is chosen such that $\dot{w}^\mu=0$ and $w^\mu w_\mu=-1$. This implies that $\dot{S}^{\mu\nu}w_\mu=0,~\ddot{S}^{\mu\nu}w_\mu=0$ and all the contractions of $w^\mu$ with the higher covariant derivatives of spin are equal to zero as well. Using this fact the contraction of \ref{eq:MPD_S} with $w_\mu$ leads to
\begin{equation}\label{eq:MomentumOKS}
    P^\mu=\frac{1}{-w_\nu \mathcal{U}^\nu}[(-P^\gamma w_\gamma) \mathcal{U}^\mu+K^{\mu\delta} w_\delta].
\end{equation}

It is obvious from \ref{eq:MomentumOKS} that the hidden momentum vanishes for \OKS~\SSC~in the pole-dipole approximation allowing the four-velocity and the four-momentum to be parallel \cite{Kyrian:2007zz}. At pole-dipole approximation relation~\ref{eq:MomentumOKS} when contracted with $\mathcal{U}_\mu$ suggests that %
\begin{align} \label{eq:MassAltDef}
    m=\frac{P^\nu w_\nu}{\mathcal{U}^\nu w_\nu}.
\end{align}
In the original work Ref.~\cite{Kyrian:2007zz}, the fact that $P^\mu=m~\mathcal{U}^\mu$ holds for \OKS~\SSC~in the pole-dipole approximation has been achieved by following a different way. In particular, in Ref.~\cite{Kyrian:2007zz} successive contractions of \ref{eq:MPD_S} 
\begin{align}
    & P^\mu \dot{S}_{\mu\nu}\dot{S}^{\nu\kappa}w_\kappa=(m^2-\mu^2)P^\kappa w_\kappa=0, \\
    & \mathcal{U}^\mu \dot{S}_{\mu\nu}\dot{S}^{\nu\kappa}w_\kappa=(m^2-\mu^2)\mathcal{U}^\kappa w_\kappa=0, 
\end{align}
led to $\mu^2=m^2$, since $\dot{S}^{\nu\kappa}w_\kappa=0$ and $w^\mu,~\mathcal{U}^\mu,~P^\mu$ are time-like vectors. Having found that $\mu^2=m^2$, the relations in \ref{eq:UPcontr} imply that $P^\mu=m~\mathcal{U}^\mu$, i.e. that the hidden momentum vanishes. The feature of a vanishing hidden momentum in the pole-dipole approximation is what makes \OKS~a very interesting \SSC. However, even if \OKS~\SSC~holds for any multipole approximation, the vanishing hidden momentum feature does not. Keeping this in mind we discuss under which condition, this feature can hold in the pole-dipole-quadrupole approximation.

In \cite{Costa:2014nta}, it has been shown that by using the hidden momentum definition on \ref{eq:MomentumOKS} leads to
\begin{align}\label{eq:MomHidOKSQ}
   P^\mu_{\rm hid} = -\frac{1}{w_\nu \mathcal{U}^\nu}(\mathcal{U}^{\mu}\mathcal{U}_\kappa+\delta^\mu_{~\kappa})K^{\kappa\gamma}w_\gamma.
\end{align}
Hence, to eliminate the hidden momentum, one needs such $w^\mu$ that 
\begin{align}\label{eq:VanHidden}
  (\mathcal{U}^{\mu}\mathcal{U}_\kappa+\delta^\mu_{~\kappa})K^{\kappa\gamma}w_\gamma=0  .
\end{align}
Note, however, that the above condition after it is contracted with $w_\mu$ leads to
\begin{align}
   w_\mu \mathcal{U}^\mu \mathcal{U}_\sigma K^{\sigma\gamma} w_\gamma=0 .
\end{align}
Since both $w_\mu$ and  $\mathcal{U}^\mu$ are time-like vectors $w_\mu\mathcal{U}^\mu\neq0$, then $\mathcal{U}_\sigma K^{\sigma\gamma} w_\gamma=0$, which in turn implies because of \ref{eq:VanHidden} that 
\begin{align}\label{eq:PDQMomVan}
    K^{\mu\gamma} w_\gamma=0.
\end{align}
Hence, the latter is actually the condition for the \OKS~\SSC~in the pole-dipole-quadrupole approximation leading to a vanishing hidden momentum. 

We can crosscheck this condition with the successive contractions of \ref{eq:MPD_S} in the pole-dipole-quadrupole approximations. These contractions read
\begin{align}
     & P^\mu \dot{S}_{\mu\nu}\dot{S}^{\nu\kappa}w_\kappa=(m^2-\mu^2)P^\kappa w_\kappa+K^{\kappa \sigma}K_{\sigma \mu}P^{\mu}w_{\kappa}+K_{\sigma \mu}\mathcal{U}^{\sigma}P^{\mu}P^{\kappa}w_{\kappa}+K^{\kappa \sigma}w_{\kappa}(\mu^2 \mathcal{U}_{\sigma}-mP_{\sigma})=0, \label{eq:OKSfixP} \\
    & \mathcal{U}^\mu \dot{S}_{\mu\nu}\dot{S}^{\nu\kappa}w_\kappa=(m^2-\mu^2)\mathcal{U}^\kappa w_\kappa+K^{\kappa \sigma}K_{\sigma \mu}\mathcal{U}^{\mu}w_{\kappa}-K_{\sigma \mu}P^{\sigma}\mathcal{U}^{\mu}\mathcal{U}^{\kappa}w_{\kappa}-K^{\kappa \sigma}(P_{\sigma}-m \mathcal{U}_{\sigma})w_{\kappa}=0. \label{eq:OKSfixU}
\end{align}
and once the condition \ref{eq:PDQMomVan} is applied, we are led to
\begin{align} \label{eq:OKSredFA}
    m^2-\mu^2+K_{\sigma\mu} \mathcal{U}^\sigma P^\mu=0 ,
\end{align}
since the contraction between time-like vectors cannot be zero.  Contracting \ref{eq:MPD_S} with $\mathcal{U}_\sigma P_\mu$ and taking into account \ref{eq:OKSredFA} results in 
\begin{align} 
    \dot{S}^{\sigma\mu}\mathcal{U}_\sigma P_\mu=0.
\end{align}
The latter has to hold for any four-velocity and four-momentum, hence $P^\mu\|\mathcal{U}^\mu$. This in turn implies through \ref{eq:OKSredFA} that $\mu^2=m^2$. The approximation independent relations \ref{eq:UPcontr} are consistent with this result, since it implies that if $\mu^2=m^2$, then $P^\mu=m~\mathcal{U}^\mu$.

Let us explore some further consequences of the condition \ref{eq:PDQMomVan}:
\begin{itemize}
    \item Since $P^\mu \| \mathcal{U}^\mu$, \ref{eq:MPD_S} is reduced to
\begin{align}\label{eq:Phid0}
    \dot{S}^{\mu\nu}=K^{\mu\nu}.
\end{align}
    \item Under condition \ref{eq:PDQMomVan}, \ref{eq:MomentumOKS} implies that \ref{eq:MassAltDef} holds also for the pole-dipole-quadrupole approximation.
    \item Since $\dot{w}^\mu=0$, then it should hold that $\dot{K}^{\mu\nu}w_\mu=0,~\ddot{K}^{\mu\nu}w_\mu=0$ and all the contractions of $w^\mu$ with the higher covariant derivatives of $K^{\mu\nu}$ are equal to zero as well, as it holds for the spin tensor. This implies a coupling of the evolution of the quadrupole moment with the derivatives of the Riemann tensor. 
\end{itemize}
Note that contrary to the \P~ and \TD~\SSC s, where the reference vectors are specific and have physical interpretation, for the \OKS~\SSC~there is a freedom of how to choose the reference vector, which should allow condition~\ref{eq:PDQMomVan} to hold for some cases. In \ref{sec:PQS} and \ref{sec:PQK} we verify for a specific quadrupole component, orbital setups and specific $w^\mu$ choices that this condition is satisfied. However, it should be stressed that these cases appear to be the exception of the rule, i.e. we do not expect condition~\ref{eq:PDQMomVan} to hold in general.

Having chosen a framework for which  $P^\mu=m \mathcal{U}^\mu$ holds, it is somehow intuitive to assume at first that $\dot{m}=0$, but this is not in general the case. In particular, contracting \ref{eq:MPD_P} with $\mathcal{U}_\mu$ and taking into account that
\begin{align}
    \dot{P}^\mu=\dot{m} \mathcal{U}^\mu+m \dot{\mathcal{U}}^\mu \label{eq:derP}
\end{align}
leads to 
\begin{align}
    \dot{m}=\dfrac{1}{6}J^{\alpha \beta \gamma \delta} \mathcal{U}_\mu \nabla^{\mu}R_{\alpha \beta \gamma \delta}. \label{eq:MassDer}
 \end{align}
Hence, only if the rhs of \ref{eq:MassDer} is zero, $m$ is a conserved quantity. In this case, since $m=\mu$, $\mu$ is conserved as well. If the mass is a constant of motion, then \ref{eq:derP} reduces to $\dot{P}^\mu=m \dot{\mathcal{U}}^\mu$ and the proportionality of $P^\mu$ and $\mathcal{U}^\mu$ holds for every higher derivatives of these vectors. Another way to address the conservation of the mass issue is to take the derivative of \ref{eq:MassAltDef}, this leads to
\begin{align}\label{eq:MassConsG}
    \dot{m}=\frac{\left(\dot{P}^\nu-m~ \dot{\mathcal{U}}^\nu\right)w_\nu}{\mathcal{U}^\nu w_\nu}.
\end{align}
In order for the above expression to be zero, either $\dot{P}^\nu=m~ \dot{\mathcal{U}}^\nu$ or the vector $\dot{P}^\nu-m~ \dot{\mathcal{U}}^\nu$ is spacelike. We have no reason to assume a priori either.

The spin measure
\begin{align}\label{eq:SpinMeas}
    S^2=\frac{1}{2}S^{\mu\nu}S_{\mu\nu}
\end{align}
is not a conserved quantity as well, since
\begin{align}\label{eq:SpinMDer}
 \dot{S^2}=S_{\mu\nu} K^{\mu\nu},   
\end{align}
where \ref{eq:Phid0} has been employed. Contrary, to the pole-dipole approximation case, for which the conservation of the mass and the spin measure depends solely on the choice of a \SSC \cite{Semerak:1999qc}, in the pole-dipole-quadrupole approximation the conservation of these quantities depends on the background geometry, see, e.g., \ref{eq:MassDer} and \ref{eq:SpinMDer}.  Actually, a background symmetry expressed by a Killing vector field $\xi^\mu$ provides a constant of motion \cite{Ehlers1977}
\begin{align}
 \mathcal{C}(\xi)=P^\mu \xi_\mu+\frac{1}{2} S^{\nu\mu}\nabla_\nu\xi_\mu. \label{eq:Cxi}
\end{align}
Note that the quadrupole moment imparts no contribution to the conserved quantities $\mathcal{C}(\xi)$.

\subsection{Kerr spacetime}

Our work concerns \S~and Kerr spacetimes. The Kerr black hole spacetime is described by an elegant multipole structure \cite{Geroch:1970cc,Geroch:1970cd,Hansen:1974zz,Thorne:1980ru}, which depends only on the mass $M$ of the black hole and the Kerr parameter $a$, i.e. its spin. The line element of a Kerr spacetime in the Boyer-Lindquist coordinates $\{t,r,\theta,\phi\}$ reads
\begin{align*}
    ds^2=g_{tt} dt^2+2 g_{t\phi} dt d\phi+g_{\phi\phi}d\phi^2+g_{rr} dr^2+g_{\theta\theta} d\theta^2\, ,
\end{align*}
where the metric coefficients are 
 \begin{align} \label{eq:KerrMetric}
   g_{tt} &=-\left(1-\frac{2 M r}{\Sigma}\right) \; ,\:
   g_{t\phi} = -\frac{2 a M r \sin^2{\theta}}{\Sigma} \; ,\:
   g_{\phi\phi} = \frac{(\varpi^4-a^2\Delta \sin^2\theta) \sin^2{\theta}}{\Sigma} \; , \nonumber\\
   g_{\theta\theta} &= \Sigma \; , \:
   g_{rr} = \frac{\Sigma}{\Delta} \; ,
 \end{align} 
with
 \begin{align}
  \Sigma = r^2+ a^2 \cos^2{\theta} \; ,\:
  \Delta = \varpi^2-2 M r \; ,\:
  \varpi^2 = r^2+a^2 \; . \label{eq:Kerrfunc} 
 \end{align}
 If we set $a=0$ in the above metric elements, we get the \S~spacetime. 

In the Kerr spacetime, we can construct an orthonormal tetrad basis, which can simplify the computations significantly \cite{Carter:1968ks}. The explicit expressions for this tetrad field are given below
\begin{eqnarray}
e^{(0)}_{\mu} & = & \left(\sqrt{\dfrac{\Delta}{\Sigma}},0,0,-a \sin^2\theta \sqrt{\dfrac{\Delta}{\Sigma}} \right), \quad
e^{(1)}_{\mu}  =  \left(0,\sqrt{\dfrac{\Sigma}{\Delta}},0,0\right) \nonumber, \\
e^{(2)}_{\mu} & = & \left(0,0,\sqrt{\Sigma},0,0 \right), \quad
e^{(3)}_{\mu}  =  \left(\dfrac{-a \sin\theta}{\sqrt{\Sigma}},0,0,\dfrac{r^2+a^2}{\sqrt{\Sigma}}\sin\theta \right).
\label{eq:Tetrad}
\end{eqnarray}
The metic tensor is related with the tetrad field through the relation $g_{\mu \nu}=e^{(\alpha)}_{\mu}e_{(\alpha)\nu}$. 

The Kerr spacetime consist of a timelike Killing field $\xi^\mu_{(t)}$ that gives through \ref{eq:Cxi} the energy
\begin{align}\label{eq:Energy}
   \mathcal{C}_t=-\left(P_t+\frac{1}{2} S^{\mu\nu} \partial_\mu g_{t\nu} \right),
\end{align}
and a space-like one $\xi^\mu_{(\phi)}$ that gives through \ref{eq:Cxi} a component of the total angular momentum
\begin{align}\label{eq:AngMom}
    \mathcal{C}_\phi=\left(P_\phi+\frac{1}{2} S^{\mu\nu} \partial_\mu g_{\phi\nu} \right),
\end{align}
for an extended test body.

\subsection{The geodesic limit} \label{sec:GeodLim}

Having introduced and discussed the framework of the \MPD~equations in the pole-dipole-quadrupole approximation under the \OKS~\SSC, let us now focus on the special cases when such an extended body follows a geodesic trajectory. For a body to follow a geodesic orbit, we set 
\begin{align}\label{eq:GoedMot}
   \dot{\mathcal{U}^{\mu}}=0 .
\end{align}
Note, that even if the latter holds, if $\dot{m}\neq 0$ and/or $\dot{P^{\mu}_{\rm hid}}\neq 0$, then $\dot{P}^{\mu} \neq 0$.\footnote{A similar phenomenon also appears in electrodynamics, and for details, we refer our readers to Refs. \cite{Costa:2012cy,Costa:2014nta,Costa:2017kdr,PhysRevD.81.104012}.} We have showed that if $w^\mu$ is chosen such that \ref{eq:PDQMomVan} holds, then $P^{\mu}_{\rm hid}=0$, so the only obstacle for $\dot{\mathcal{U}}^\mu=0$ to imply $\dot{P}^\mu = 0$ is that one needs $\dot{m} = 0$.

In the special case that $\dot{m}=0$, given that the particle is following a geodesic on a Kerr spacetime, for which it holds that $\mathcal{U}_t$ and $\mathcal{U}_\phi$ are constants of motion, then $P_t$ and $P_\phi$ are constants as well. This in turn implies, via \ref{eq:Energy} and \ref{eq:AngMom}, that $\frac{1}{2} S^{\mu\nu} \partial_\mu g_{t\nu}$ and $\frac{1}{2} S^{\mu\nu} \partial_\mu g_{\phi\nu}$ are conserved too. Therefore, to sum up:
\begin{eqnarray}
E_{\rm g}&=&-P_t =  \text{constant},\: E_{\rm s}=- \frac{1}{2} S^{\mu\nu} \partial_\mu g_{t\nu}=\text{constant}, \nonumber \\
L_z&=&P_\phi=\text{constant}, \: J_{\rm s} = \frac{1}{2} S^{\mu\nu} \partial_\mu g_{\phi\nu}=\text{constant}.
\label{eq:conserve_geo_spin}
\end{eqnarray}
Note that the contributions from the dipole moment are only contained in the terms $E_{\rm s}$ and $J_{\rm s}$. 

With this, we finish our discussion on the underlying formalism and the key equations. In the following sections, we employ the above discussed framework in some examples to illustrate various possibilities related to the geodesic limit (\ref{eq:GoedMot}) of an extended body. In these scenarios, whenever it is needed to consider the quadrupole moment, we restrict the analysis to the mass quadrupole moment, for which the quadrupole tensor~\ref{eq:QuadTens} reduces to
\begin{equation}\label{eq:QuadMassOKS}
J^{\alpha \beta \gamma \delta}=-3 w^{[\alpha}Q^{\beta][\gamma}w^{\delta]}=-\dfrac{3}{4}\Bigl(w^{\alpha}Q^{\beta \gamma}w^{\delta}-w^{\beta}Q^{\alpha \gamma}w^{\delta}-w^{\alpha}Q^{\beta \delta}w^{\gamma}+w^{\beta}Q^{\alpha \delta}w^{\gamma}\Bigr),
\end{equation}
and in particular to the spin-induced quadrupole model case, for which
\begin{align}\label{eq:SpinInd}
    Q^{\beta\gamma}=C_{S^2}{S^\beta}_\alpha S^{\alpha\gamma},
\end{align}
where $C_{S^2}$ is a constant depending on the internal structure of the body \cite{Steinhoff:2012rw}.

\section{Geodesic trajectories of an extended body\newline in the \S~background}\label{sec:SchwBH}

We start our examples with the \S~spacetime, for which the non-vanishing metric coefficient \ref{eq:KerrMetric} reduce to
 \begin{align} \label{eq:SMetric}
   g_{tt} &=-f=-\left(1-2M/r\right) \; ,\:
   g_{\phi\phi} = r^2 \sin^2\theta \; , \:
   g_{\theta\theta} = r^2 \; , \:
   g_{rr} = f^{-1} \; .
 \end{align}

\subsection{Pole-dipole body}\label{sec:PDS}
The equations of motion for a pole-dipole test body are given by  \ref{eq:MPD_P} and \ref{eq:MPD_S} when $J^{\alpha \beta \gamma \delta}=0$. In order to have an extended body following a geodesic trajectory, we need to ensure that both \MPD~and geodesic equations are simultaneously satisfied. For a geodesic orbit in \S~spacetime we have 
\begin{align}
\mathcal{U}_{t}& =-\tilde{E}_{\rm g}, \quad \mathcal{U}_{r}=\pm \left\{\tilde{E}_{\rm g}^2-f\left(1+\frac{\tilde{L}^2_{\rm z}}{r^2}\right)\right\}^{1/2} f^{-1},\nonumber \\
\mathcal{U}_\theta&= 0, \quad \mathcal{U}_{\phi}=\tilde{L}_{\rm z},
\label{eq:4-velocity_02}
\end{align}
where $\tilde{E}_{\rm g}$ and $\tilde{L}_{\rm z}$ denote the conserved specific energy and orbital angular momentum. Without loss of generality, because of the spherical symmetry, we have set for simplicity $\mathcal{U}^{\theta}=0$ to constrain the motion on the equatorial plane. 

Since, for a pole-dipole body under \OKS~\SSC~ $\dot{m}=0$, \ref{eq:GoedMot} is equivalent to $\dot{P}^\mu=0$, we arrive to 
\begin{eqnarray}
\dot{P}^{t} &=& \dfrac{M}{r^3} \Bigl\{2 S^{tr}\mathcal{U}_{r}- S^{t\phi}\mathcal{U}_{\phi}\Bigr\}=0, \quad \dot{P}^{r} = -\dfrac{M}{r^3}\Big\{2 S^{tr}\mathcal{U}_{r}-2 S^{\phi r}\mathcal{U}_{\phi}\Bigr\}=0,\nonumber \\
\dot{P}^{\theta} &=& \dfrac{M}{r^3} \Bigl\{-2 S^{\theta \phi}\mathcal{U}_{\phi}-S^{\theta r}\mathcal{U}_{r}-S^{\theta t}\mathcal{U}_{t}\Bigr\}=0, \quad \dot{P}^{\phi} = -\dfrac{M}{r^3}\Bigl\{S^{\phi t} \mathcal{U}_{t}+S^{\phi r}\mathcal{U}_{r}\Bigr\}=0.
\label{eq:sch_pole_dipole}
\end{eqnarray}
Rearranging the above equations leads to
\begin{equation}
S^{tr}=\dfrac{S^{t\phi}\tilde{L}_{\rm z}}{2 \mathcal{U}_{r}}, \quad S^{t\theta}=-\dfrac{S^{\theta r}\mathcal{U}_{r}+2 S^{\phi \theta}\tilde{L}_{\rm z}}{\tilde{E}_{\rm g}}, \quad S^{t\phi}=-\dfrac{S^{\phi r}\mathcal{U}_{r}}{\tilde{E}_{\rm g}}.
\label{eq:spin_SBH}
\end{equation}
There is no fourth equation, because from \ref{eq:sch_pole_dipole} only three of the four equations are linearly independent. Having constrained the motion of a spinning body on the equatorial plane one would expect that $S^{\theta \mu}=0$ \cite{Harms2016}, however, when the spinning body is following a geodesic the spin should not be able to affect the motion, since the spin curvature coupling is annihilated \ref{eq:sch_pole_dipole}, hence the $S^{\theta \mu}$ components are not necessarily zero. Taking into account \ref{eq:Energy}, \ref{eq:AngMom}, \ref{eq:conserve_geo_spin}, \ref{eq:spin_SBH},  the conserved quantities can be rewritten as 
\begin{equation}\label{eq:SchConS}
\mathcal{C}_{t}=-P_t-\dfrac{L_zMS^{t\phi}}{2r^2 \mathcal{U}_{r}}, \qquad E_{\rm s}=-\dfrac{L_zMS^{t\phi}}{2r^2 \mathcal{U}_{r}}, \qquad \mathcal{C}_{\phi}=P_{\phi}-r S^{\phi r}, \qquad J_{\rm s}= -r S^{\phi r}.
\end{equation}
Since the \S~spacetime is spherical symmetric, there are two extra constants of motion
\begin{align}
        \mathcal{C}_x & = -\sin\phi\: P_\theta - \cos\phi\: \cot\theta\: P_\phi + r^2\cos\phi\: \sin^2\!\theta\: S^{\theta\phi} + r\sin\phi\: S^{\theta r} + r\cos\phi\: \sin\theta\: \cos\theta\: S^{\phi r} \:, \\
        \mathcal{C}_y & = \cos\phi\: P_\theta - \sin\phi\: \cot\theta\: P_\phi + r^2\sin\phi\: \sin^2\!\theta\: S^{\theta\phi} - r\cos\phi\: S^{\theta r} + r\sin\phi\: \sin\theta\: \cos\theta\: S^{\phi r} \:,
\end{align}
which on the equatorial plane for the geodesic motion reduce to
\begin{align}\label{eq:CxCyRed}
        \mathcal{C}_x & =  r^2\cos\phi\:  S^{\theta\phi} + r\sin\phi\: S^{\theta r} , \nonumber \\
        \mathcal{C}_y & =  r^2\sin\phi\:  S^{\theta\phi} - r\cos\phi\: S^{\theta r}.
\end{align}

Having enforced geodesic motion, let us now see what are the consequences on the conserved quantities $E_{\rm s}$, $J_{\rm s}$, $\mathcal{C}_x$  and $\mathcal{C}_y$ involving spin contributions. The conservation of $E_{\rm s}$ implies that
\begin{equation}
\dfrac{dE_{\rm s}}{d\tau}=-\dfrac{M L_{\rm z}}{2} \Bigl\{\dfrac{1}{r^2 \mathcal{U}_{r}}\dfrac{dS^{t\phi}}{d\tau}-\dfrac{2 S^{t\phi} \mathcal{U}^{r}}{r^3 \mathcal{U}_{r}}-\dfrac{S^{t\phi}}{r^2 \mathcal{U}^2_r}\dfrac{d\mathcal{U}_{r}}{d\tau}\Bigr\}=0,
\end{equation}
which can be expressed by using the parallel transport of the spin tensor evolution and the geodesic equation as follows
\begin{equation}
\dfrac{M L_{\rm z}}{2} \Bigl\{\dfrac{1}{r^2 \mathcal{U}_{r}}\left[-\left(\Gamma^{t}_{\alpha \beta}S^{\alpha \phi}\mathcal{U}^{\beta}+\Gamma^{\phi}_{\alpha \beta}S^{t \alpha}\mathcal{U}^{\beta}\right)\right]-\dfrac{2 S^{t\phi}}{r^3 g_{rr}}-\dfrac{S^{t\phi}}{r^2 \mathcal{U}^2_r}\left(\Gamma^{\alpha}_{r\beta}\mathcal{U}^{\beta}\mathcal{U}_{\alpha}\right)\Bigr\}=0.
\end{equation}
By expanding the above equation, we reach
\begin{align}
\dfrac{M L_{\rm z}S^{\phi r}}{4(r-2M)^2r^5 \tilde{E}_{\rm g}\mathcal{U}_{r}}\Bigl\{L^2_{z}(r^2-8Mr+12M^2)-2r^2(r^2-6Mr)(\tilde{E}^2_{\rm g}-1)+16M^2 r^2\Bigr\}=0.
\end{align}
For an equatorial orbit the above equation can be zero if $S^{\phi r}=0$ or $L_{\rm z}=0$, or the expression in the curl brackets is zero. The latter would imply that a spinning body can follow circular an equatorial geodesic orbit. 

Let us breakdown these three cases to see whether they are possible and where do they lead to.
\begin{itemize}
 \item If we assume that there is a circular orbit, then, since $\mathcal{U}^r=0$, \ref{eq:spin_SBH} implies that  $S^{t\phi}=0$. By substituting this in the $\dot{S}^{t\phi}=0$ equation, we find that $E=L_{\rm z}=0$ is needed. However, such a solution is inconsistent with the $\mathcal{U}^\nu \mathcal{U}_\nu=-1$ constraint. Hence, a spinning body under \OKS~\SSC~ moving in the Schwarzschild background cannot follow a geodesic circular orbit.
 \item If we assume that $S^{\phi r}=0$ and $L_{\rm z}\neq 0$, then \ref{eq:spin_SBH} implies that both $S^{tr}$ and $S^{t\phi}$ vanish as well. This leads in turn from \ref{eq:SchConS} that $E_{\rm s}=J_{\rm s}=0$. If we combine the expansions of the $\dot{S}^{\theta \mu}=0$ equations with the middle relation of \ref{eq:spin_SBH}, then we arrive to $L_{\rm z}(S^{\theta r} \mathcal{U}^\phi+S^{\theta \phi} \mathcal{U}^r)=0$, which is true if $S^{\theta \mu}=0$, but then the body is left without spin. If the spin tensor components are not zero, then after some further manipulations of the $\dot{S}^{\theta \mu}=0$ equations, during which we take into account that $S^{\theta r} \mathcal{U}^\phi+S^{\theta \phi} \mathcal{U}^r=0$, we end up with $\displaystyle \frac{d \mathcal{U}^r}{d \tau}=-\frac{L_{\rm z}^2}{r^3}$. This result is inconsistent with the geodesic equations. Hence, $S^{\phi r}=0$ option cannot provide a geodesic orbit for a spinning body under \OKS~\SSC.
 \item For $L_{\rm z}=0$, \ref{eq:SchConS} implies that $E_{\rm s}=0$ and \ref{eq:spin_SBH} implies that $S^{tr}=0$. It can be shown
 \begin{itemize}
     \item by expressing $\dot{S}^{\theta r}=0$ in terms of Christoffel symbols that $r\,S^{r \theta}= \mathcal{K}_r=\textrm{constant}$.
     \item by expressing $\dot{S}^{\theta \phi}=0$ in terms of Christoffel symbols that $r^2 S^{\theta \phi}=\mathcal{K}_\phi=\textrm{constant}$.     
     \item from \ref{eq:CxCyRed} that
      \begin{align}
        \mathcal{C}_x &  =  \cos\phi\mathcal{K}_{\phi} - \sin\phi \mathcal{K}_r , \nonumber \\
        \mathcal{C}_y & =  \sin\phi\:  \mathcal{K}_{\phi} + \cos\phi\: \mathcal{K}_{r}.
     \end{align}
     Hence, for the radial motion during which $\phi$ is constant, $\mathcal{K}_r$ and $\mathcal{K}_\phi$ are recast expressions of  $\mathcal{C}_x$ and $\mathcal{C}_y$.
     \item by expressing $\dot{S}^{r \phi}=0$ in terms of Christoffel symbols that $r^2 S^{r \phi}=\textrm{constant}$, which is just reproducing $J_s$.     
     \item from the middle relation of \ref{eq:spin_SBH} that $S^{\theta t}=-(\mathcal{K}_r \mathcal{U}_r)/(r \tilde{E}_{\rm g})$.
     \item by expressing $\dot{S}^{tr}$, $\dot{S}^{\theta t}$, $\dot{S}^{\phi t}$ and $\dot{S}^{t r}$ in terms of Christoffel symbols that they just trivially vanish.
 \end{itemize} 
Note that $J_s$, $\mathcal{C}_x$ and $\mathcal{C}_y$ reemerged naturally, so they are consistent with the radial geodesic motion of a spinning body. Moreover, since $S^{\phi t}$ can also be expressed from the third relation in \ref{eq:sch_pole_dipole}, the spin tensor takes the following form
\begin{align}
S^{\mu \nu} &=\left({\begin{array}{cccc} 0 & S^{tr} & S^{t\theta} & S^{t\phi} \\
-S^{tr} & 0 & S^{r\theta} & S^{r\phi} \\
-S^{t\theta} & -S^{r\theta} & 0 & S^{\theta \phi}\\
-S^{t\phi} & -S^{r \phi} & -S^{\theta \phi} & 0 \end{array}}\right) \nonumber \\
&= \left({\begin{array}{cccc} 0 & 0 & {(\mathcal{K}_r \mathcal{U}_r)}/(r \tilde{E}_{\rm g}) & {(J_{\rm s}~\mathcal{U}_r)}/(r \tilde{E}_{\rm g}) \\
0 & 0 & {\mathcal{K}_r}/{r} & {J_{\rm s}}/{r} \\
-{(\mathcal{K}_r \mathcal{U}_r)}/(r \tilde{E}_{\rm g}) & -{\mathcal{K}_r}/{r} & 0 & \mathcal{K}_{\phi}/r^2\\
-{(J_{\rm s}~\mathcal{U}_r)}/(r \tilde{E}_{\rm g}) & -{J_{\rm s}}/{r}  & -\mathcal{K}_{\phi}/r^2 & 0 \end{array}}\right), \label{eq:SPIN_TENSOR_FORM}
\end{align} 
which shows that after fixing the initial conditions the spin tensor is just function of $r$, since $\mathcal{U}_r(r,\tilde{E}_{\rm g},J_{\rm s})$ (\ref{eq:4-velocity_02}). In conclusion, $L_{\rm z}=0$ corresponds to a consistent solution of the \MPD~equations, which shows a spinning body falling radially into a \S~black hole following a geodesic trajectory. An interesting outcome of the above calculation is that the condition $S^{\alpha \beta}\mathcal{U}_{\beta}=0$, which is the \P~\SSC, naturally emerges from \ref{eq:spin_SBH}. Furthermore, since for \OKS~\SSC~it holds that $P^\mu||\mathcal{U}^\mu$, we have that the centroids of the \OKS, \P~and \TD~($S^{\mu\nu}P_\mu=0$) \SSC s coincide. This fact was already noticed in \cite{Costa:2017kdr}, and the corresponding trajectory is known in the literature  \cite{Costa:2012cy,Costa:2014nta,Costa:2017kdr}. The novel result is that the radial motion is the only geodesic solution for a spinning body under \OKS~\SSC~in the \S~spacetime. 
\end{itemize}
 
\subsection{Pole-dipole-quadrupole body} \label{sec:PQS}
Let us now examine whether a pole-dipole quadrupole body under \OKS~\SSC~can follow a geodesic trajectory. Since the geodesic motion on \S~background is equatorial, i.e. $\mathcal{U}_{\theta}=0$, the parallel transport of the $\theta$ component of the vector $w^\mu$ reduces to
\begin{equation}\label{eq:wth_par}
\dfrac{dw^{\theta}}{d\tau}+\dfrac{\mathcal{U}^{r}}{r}w^{\theta}=0,  
\end{equation}
which translates to $r w^{\theta}=\text{constant}$. We assume for simplicity that this constant is zero, hence $w^{\theta}=0$.

Our investigation focuses on a solely spin induced quadrupole moment given by \ref{eq:SpinInd}. Since in \ref{sec:PDS} we have shown that only a radially moving pole-dipole body in a \S~background can follow a geodesic orbit, it would be interesting to see if this is also the case when the spin induced quadrupole moment is added to the model.  Hence, we study only the radial geodesic \ref{eq:4-velocity_02} and check whether it can be a solution to the \MPD~equations, as in the pole-dipole case (\ref{sec:PDS}). On this radial orbit, $w_{\phi}$ follows an equation similar to \ref{eq:wth_par}, which allows us to assume again for simplicity that $w^{\phi}=0$. The other components of the reference vector $w_\mu$ read
\begin{eqnarray} \label{eq:refvect_SchRad}
    \dfrac{dw_t}{d\tau}&=&\dfrac{M}{r^2}\Big(\mathcal{U}_rw_t-\mathcal{U}_t w_r\Big), \nonumber \\
 \dfrac{dw_r}{d\tau}&=& -M \Big[\dfrac{\mathcal{U}_t w_t}{(r-2M)^2}+\dfrac{\mathcal{U}_r w_r}{r^2}\Big].
\end{eqnarray}
Note that \ref{eq:4-velocity_02} and \ref{eq:refvect_SchRad} lead to $\dfrac{d}{d\tau}(-\mathcal{U}_t w_r+\mathcal{U}_{r}w_t)=0$, which implies that along the radial geodesic trajectory $-\mathcal{U}_t w_r+\mathcal{U}_{r}w_t=-C_r={\rm constant}$. 

From the \SSC~equations and by using
\begin{align} \label{eq:w_rad}
   w^{\theta}=w^{\phi}=0,
\end{align}
we obtain
\begin{eqnarray}
 S^{tr}=0, \quad S^{r\theta}=-S^{t\theta}w_t/w_r, \quad S^{r\phi}=-S^{t\phi}w_t/w_r. \label{eq:Scomp}
\end{eqnarray}
When \ref{eq:w_rad} and \ref{eq:Scomp} are implemented on $K^{\mu\nu}$ leads to the following non-diagonal components of the tensor:
%%%%
\begin{eqnarray}
&&    K^{tr}=0,\quad K^{t\theta}=\dfrac{3C_{S^2}MS^{r\phi}S^{\theta \phi}w_r}{rw_t}, \quad K^{t\phi}=-\dfrac{3C_{S^2}MS^{r\theta}S^{\theta \phi}w_r}{rw_t}, \nonumber \\
&&  K^{r\theta}=-\dfrac{3C_{S^2}MS^{r\phi}S^{\theta \phi}}{r}, \quad K^{r\phi}=\dfrac{3C_{S^2}MS^{r\theta}S^{\theta \phi}}{r}.
\end{eqnarray}
Using the above expressions, it is straightforward to see that all the four components of $K^{\mu\nu} w_\nu$ identically vanish. Hence, the hidden momentum vanishes, i.e. $P^\mu=m \mathcal{U}^\mu$, on a radial trajectory of the under investigation extended body.

Having verified that the hidden momentum vanishes, we can now investigate \ref{eq:MPD_P}, which can be rewritten as
\begin{equation}\label{eq:MPD_PG}
    \dot{m}\mathcal{U}^{\mu}+\frac{1}{2}R^{\mu}_{~\nu \alpha \beta}\mathcal{U}^{\nu} S^{\alpha \beta}+\dfrac{1}{6}J^{\alpha \beta \gamma \delta} \nabla^{\mu}R_{\alpha \beta \gamma \delta}=0,
\end{equation}
since for a geodesic orbit $\dot{\mathcal{U}}^{\mu}=0$. 

\paragraph{Time component:}

For the time component $\mu=t$, \ref{eq:MPD_PG} reduces to $\dot{m}=0$, since using \ref{eq:w_rad}, \ref{eq:Scomp} and being on a radial geodesic leads to $R^{t}_{~\nu \alpha \beta}\mathcal{U}^{\nu} S^{\alpha \beta}=0$ and $J^{\alpha \beta \gamma \delta} \nabla^{t}R_{\alpha \beta \gamma \delta}=0$ and $\mathcal{U}^t \neq 0$. When we expand \ref{eq:MassDer} and recall that $w^{\mu}w_{\mu}=-1$, we obtain
\begin{equation}
    \dot{m}=\dfrac{3MC_{S^2}(r-2M)\mathcal{U}_r}{2r^3w^2_r}\left\{[(S^{t\theta})^2+(S^{t\phi})^2]-2 r^2 w^2_r(S^{\theta \phi})^2\right\}=0. \label{eq:mdot}
\end{equation}
This implies that $(S^{t\theta})^2+(S^{t\phi})^2=2r^2w^2_r(S^{\theta \phi})^2$. For the remaining three components three components of \ref{eq:MPD_PG} we take into account that $\dot{m}=0$.

\paragraph{Radial component:}
For the radial component $\mu=r$, \ref{eq:MPD_PG} reads 
\begin{equation} \label{eq:SpinIndCon}
    \dfrac{3C_{S^2}M(r-2M)}{2r^3w^2_r}\left\{[(S^{t\theta})^2+(S^{t\phi})^2]-2r^2 (S^{\theta \phi})^2 w^2_r\right\}=0.
\end{equation}
Note that \ref{eq:SpinIndCon} and \ref{eq:mdot} are referring to the same set of constraints on the spin tensor.

\paragraph{Polar component:} For $\mu=\theta$ component, \ref{eq:MPD_PG} provides the relation 
\begin{equation} 
\dfrac{MS^{t\theta}C_r}{r w_r}+3C_{S^2}MS^{t\phi}S^{\theta \phi}w_t=0. \label{eq:theta_EOM}
\end{equation}

\paragraph{Azimuthal component:}
For $\mu=\phi$ component \ref{eq:MPD_PG} provides the relation 
\begin{equation}
 \dfrac{MS^{t\phi}C_r}{r w_r}-3C_{S^2}MS^{t\theta}S^{\theta \phi}w_t=0. \label{eq:phi_EOM}   
\end{equation}
In \ref{eq:theta_EOM} and \ref{eq:phi_EOM} there are both the dipole and the quadrupole contribution assuming that $C_r\neq 0$ and $C_{S^2}\neq 0$. Having done this assumption, \ref{eq:phi_EOM} can be written as 
\begin{align}
   S^{t\phi}=\dfrac{3C_{S^2}S^{t\theta}S^{\theta \phi}r w_t w_r}{C_r}. \label{eq:step1}
\end{align}
Plugging this expression in~\ref{eq:theta_EOM}, we obtain
$$S^{t\theta}\left[\left(\dfrac{C_r}{r w_r}\right)^2+(3C_{S^2} S^{\theta \phi}w_t)^2\right]=0.$$
The latter leads to $S^{t\theta}=0$, since we have assumed that $C_r\neq0$. In turn, \ref{eq:step1} leads to $S^{t\phi}=0$ and \ref{eq:SpinIndCon} to $S^{\theta \phi}=0$. Moreover, from \ref{eq:Scomp}, we notice that all the other spin components vanish as well, i.e., $S^{tr}=S^{r\theta}=S^{r\phi}=0$. Hence, the body is not spinning. 

If $C_r=0$, then \ref{eq:theta_EOM} and \ref{eq:phi_EOM} are simultaneously satisfied if $S^{\theta\phi}=0$, which because of \ref{eq:SpinIndCon} and \ref{eq:Scomp} finally leads again to $S^{\mu\nu}=0$. Hence, $C_r=0$ is not a sufficient condition to lead to a vanishing spin. Note, that the $C_r=0$ case implies that $\mathcal{U}^\mu=w^\mu$. In order to show this, recall that both $\mathcal{U}^\mu$ and $w^\mu$ are timelike future pointing unit vectors and they have only time and radial components. Hence, $C_r=0$ leads to ${w_r}^2={\mathcal{U}_r}^2$, which in turn implies that $C_r=0 \Leftrightarrow \mathcal{U}^{[\mu} w^{\nu ]} = 0$ gives $w^\mu=\mathcal{U}^\mu$. We see now that for $C_r=0$ our discussion includes also the \P~\SSC~and the \TD~\SSC. Moreover, since $C_r=0$ implies also a vanishing dipole contribution to the acceleration, while the quadrupole contribution vanishes independently, it is important to mention that this coincidence between the centroids is comprised in the \ref{eq:spin_SBH} obtained already in~\ref{sec:PDS}.

Therefore, for $w^\theta=w^\phi=0$ we are led to the conclusion: \textit{a pole-dipole-(spin induced) quadrupole body under \OKS~\SSC~cannot follow a radial geodesic trajectory}. Note that we have not examined the cases for which $w^\theta={\rm constant}_1/r$ and $w^\phi={\rm constant}_2/r$. However, for continuity reasons with respect to ${\rm constant}_i (i=1,2)$, we would not expect this to change our conclusion. 

\section{Geodesic trajectories of an extended body in the Kerr background}\label{sec:KerrBH}

The results we present for the Kerr background in this section reduce for $a=0$ to the \S~case of \ref{sec:SchwBH}. However, we find that since the \S~case is simpler than Kerr, one can gain more insight into the results discussed in our work. Therefore, we have presented the \S~case first, before discussing the more complicated Kerr case.  

\subsection{Pole-dipole body}\label{sec:PDK}

When dealing with the pole-dipole approximation in Kerr, we have found that the \MPD~equations simplify significantly when the tetrad field~\ref{eq:Tetrad} is employed. However, this simplification is not sufficient to allow us to deal with generic orbits. Hence, we have confined our study on the equatorial plane by taking advantage of the fact that since $\mathcal{U}^{(\theta)}=e^{(2)}_{2}\mathcal{U}^{\theta}$ when $\mathcal{U}^{\theta}=0$, then $\mathcal{U}^{(\theta)}=0$ as well. Since $\dot{m}=0$ in the pole-dipole approximation, assuming geodesic motion $\dot{\mathcal{U}}^{(\nu)}=0$ implies that \ref{eq:MPD_P} reduces to
\begin{eqnarray}
\dot{P}^{(t)} &=& \dfrac{M}{r^3}\left\{2S^{(t)(r)}\mathcal{U}^{(r)}-S^{(t)(\phi)}\mathcal{U}^{(\phi)}\right\}=0, \nonumber \\
\dot{P}^{(r)} &=& \dfrac{M}{r^3}\left\{2S^{(t)(r)}\mathcal{U}^{(t)}-S^{(r)(\phi)}\mathcal{U}^{(\phi)}\right\}=0, \nonumber \\
\dot{P}^{(\theta)} &=& \dfrac{M}{r^3}\left\{2S^{(\theta)(\phi)}\mathcal{U}^{(\phi)}+S^{(r)(\theta)}\mathcal{U}^{(r)}-S^{(t)(\theta)}\mathcal{U}^{(t)}\right\}=0,\nonumber \\
\dot{P}^{(\phi)} &=& \dfrac{M}{r^3}\left\{S^{(r)(\phi)}\mathcal{U}^{(r)}-S^{(t)(\phi)}\mathcal{U}^{(t)}\right\}=0, 
\end{eqnarray}
From the first three expressions of above set of equations we get:
\begin{equation}
S^{(t)(r)}=\dfrac{S^{(t)(\phi)}\mathcal{U}^{(\phi)}}{2\mathcal{U}^{(r)}}, \quad S^{(t)(\phi)}=\dfrac{S^{(r)(\phi)}\mathcal{U}^{(r)}}{\mathcal{U}^{(t)}}, \quad S^{(t)(\theta)}=\dfrac{S^{(r)(\theta)}\mathcal{U}^{(r)}+2S^{(\theta)( \phi)}\mathcal{U}^{(\phi)}}{\mathcal{U}^{(t)}},
\label{eq:Spin_Rel}
\end{equation}
which is similar to that of the \S~case (\ref{eq:spin_SBH}). As in that case the fourth equation is trivially satisfied because of the middle relation of \ref{eq:Spin_Rel}. 

In the tetrad field frame (\ref{eq:Tetrad}), the conserved quantities given in \ref{eq:conserve_geo_spin} read
\begin{equation}\label{eq:PDConQuaK}
%    \mathcal{C}_{t}=-P_t-\dfrac{MS^{(t)(\phi)}\mathcal{U}^{(\phi)}}{2r^2 \mathcal{U}^{(r)} },\quad
E_{\rm s}=-\dfrac{MS^{(t)(\phi)}\mathcal{U}^{(\phi)}}{2r^2 \mathcal{U}^{(r)}}, \quad
%\mathcal{C}_{\phi}=P_{\phi}-\dfrac{\sqrt{\Delta}}{r}S^{(r)(\phi)}, \quad 
J_{\rm s}=-\dfrac{\sqrt{\Delta}}{r}S^{(r)(\phi)}.
\end{equation}
As in \ref{sec:PDS}, we will check the consequences of having enforced the geodesic motion on the extended body by checking the above conserved quantities $E_{\rm s},~J_{\rm s}$. Let us start from
\begin{equation}\label{eq:derEsK}
    \frac{d E_{\rm s}}{d\tau}=\dfrac{3MS^{(r)(\phi)}\mathcal{U}^{(r)}\mathcal{U}^{(\phi)}}{2r^4(\mathcal{U}^{(t)})^2}\left\{2\mathcal{U}^{(t)}\sqrt{\Delta}+a\mathcal{U}^{(\phi)}\right\}=0,
\end{equation}
which implies the following cases:
\begin{itemize}

    \item If~~$\mathcal{U}^{(r)}=0$, then the body has to follow circular geodesic orbits. By using \ref{eq:Spin_Rel}, this leads to $S^{(t)(r)}=S^{(t)(\phi)}=0$. If we now expand the expression $\dot{S}^{(t)(\phi)}=0$,   we get $(a^2-Mr)\mathcal{U}^{(t)}+a\sqrt{\Delta}\mathcal{U}^{(\phi)}=0$. Substituting this expression in the timelike condition, i.e., $[\mathcal{U}^{(t)}]^2-[\mathcal{U}^{(\phi)}]^2=1$, we obtain, $[\mathcal{U}^{(t)}]^2=\dfrac{a^2 \Delta}{r^2(a^2-M^2)}$. Since $\Delta/(a^2-M^2)$ is always negative outside the event horizon for a Kerr black hole, the case $\mathcal{U}^{(r)}=0$ does not hold.  
    
    \item In the case that the quantity in the bracket is equal to zero, since $\mathcal{U}^{(t)},~\mathcal{U}^{(\phi)}$ are functions of the radius and the constants $E_g,~L_z$, the body should be on a circular orbit of constant radius. A circular orbit implies again that $\mathcal{U}^{(r)}=0$. Hence, this case cannot hold.  
    \item For the case that $S^{(r)(\phi)}=0$, we get $S^{(t)(r)}=S^{(t)(\phi)}=0$ from \ref{eq:Spin_Rel}. However, if we differentiate the third expression in \ref{eq:Spin_Rel} with respect to $\tau$, we arrive at 
    \begin{equation}\label{eq:Difeq:Spin_Rel}
        \dfrac{3\mathcal{U}^{(\phi)}}{r^2}\bigl\{a(S^{(r)(\theta)}\mathcal{U}^{(t)}-S^{(t)(\theta)}\mathcal{U}^{(r)}]+\sqrt{\Delta}[S^{(r)(\theta)}\mathcal{U}^{(\phi)}-S^{(\theta)( \phi)}\mathcal{U}^{(r)}]\bigr\}=0,
    \end{equation}
    by using the fact that $\dot{S}^{(\mu) (\nu)}=0$, $\dot{\mathcal{U}}^{(\mu)}=0$ and their expansions into Christoffel symbols and the spin connection. By expanding $\dot{S}^{(\theta)(\phi)}=0$, one obtains $rS^{(\theta)( \phi)}=\text{constant}$. Combining the latter result, \ref{eq:Difeq:Spin_Rel}, \ref{eq:Spin_Rel} and the fact that $\mathcal{U}^{(t)},~\mathcal{U}^{(r)},~\mathcal{U}^{(\phi)}$ are functions of the radius and the constants $E_g,~L_z$ allows us to express all the non-vanishing spin components in terms of the radial coordinate $r$ and the constants $E_g,~L_z$. However, plunging these expression of the spin components into $\dot{S}^{(\mu)(\nu)}=0$ one runs into inconsistencies with the geodesic equation, on that ground, this case is excluded. 
    
    \item If $\mathcal{U}^{(\phi)}=0$, then it is immediately implied that $L_{\rm z}=aE_{\rm g}$. From \ref{eq:PDConQuaK} we see that $E_{\rm s}=0$, while from \ref{eq:Spin_Rel} we get $S^{(t)(r)}=0$.  Moreover, it can be shown that
    \begin{itemize}
        \item the expansion of $\dot{S}^{(r)(\phi)}=0$ leads to $\sqrt{\Delta}S^{(r)(\phi)}/r=\text{constant}$, which is actually $J_{\rm s}$.  Interestingly, this expression can also be written as $S^{(r)(\phi)}/\mathcal{U}^{(t)}=\text{constant}$, as we have $\mathcal{U}^{(t)}=E_{\rm g}r/\sqrt{\Delta}$.
        
        \item the expansion of $\dot{S}^{(t)(\phi)}=0$, leads to $S^{(t)(\phi)}/\mathcal{U}^{(r)}=\text{constant}$, which is in agreement with the middle relation in \ref{eq:Spin_Rel} when we take into account that  $S^{(r)(\phi)}/\mathcal{U}^{(t)}=\text{constant}$.
        
        \item $\dot{S}^{(t)(r)}=0$ is trivially satisfied. 
        
        \item by using the relations $\dot{S}^{(\theta)(\mu)}=0$, we arrive at the following expressions:
        \begin{eqnarray} \label{eq:PDKradS} \dfrac{d}{d\tau}(S^{(r)(\theta)}/\mathcal{U}^{(t)})=-(a/r^2)S^{(\theta)(\phi)}, \nonumber \\
 \dfrac{d}{d\tau}(S^{(t)(\theta)}/\mathcal{U}^{(r)})=-(a/r^2)S^{(\theta)(\phi)}, \nonumber \\
   \dfrac{dS^{(\theta )(\phi)}}{d\tau}=(a/r^2)(S^{(r)(\theta)}/\mathcal{U}^{(t)}).
        \end{eqnarray}
        It is interesting to note that the subtraction of the first two evolution equations in \ref{eq:PDKradS} leads to a conserved quantity, which in fact can be retrieved by the right relation of \ref{eq:Spin_Rel} for $\mathcal{U}^{(\phi)}=0$. This quantity reads        $$S^{(r)(\theta)}\mathcal{U}^{(r)}-S^{(t)(\theta)}\mathcal{U}^{(t)}=0.$$
        Note that in fact \ref{eq:Spin_Rel} for $\mathcal{U}^{(\phi)}=0$ implies that the \P~\SSC~holds along with \OKS, which in turn implies that \TD~\SSC~is satisfied as well. To conclude, we have not only found a solution providing geodesic orbits for a spinning particle in Kerr background, but we have also shown that this solution is the only possible for equatorial orbits in Kerr under \OKS~\SSC. 
    \end{itemize}

\end{itemize}

\subsection{Pole-dipole-quadrupole body} \label{sec:PQK}

Let us study now a pole-dipole-quadrupole body with spin induced quadrupole in a Kerr spacetime by following a similar route as in the \S~BH case (\ref{sec:PQS}). We assume that the particle follows the geodesic trajectory that we found in \ref{sec:PDK} to be compatible with a pole-dipole body. For this trajectory holds that $L_{\rm z}=aE_{\rm g}$ and the respective four-velocity reads:
\begin{align}
\mathcal{U}_{t} =-\tilde{E}_{\rm g}, \quad \mathcal{U}_{r}=r^2\mathcal{U}^r \Delta^{-1}=\pm r \left\{\tilde{E}^2_{\rm g}r^2- \Delta\right\}^{1/2} \Delta^{-1},\quad
\mathcal{U}_\theta= 0, \quad \mathcal{U}_{\phi}=\tilde{L}_{\rm z},
\label{eq:4-velocity_03}
\end{align}

Regarding the reference vector $w^{\mu}$, note that, as in the Schwarzschild case, for the equatorial motion in Kerr it holds that $w^\theta r={\rm constant}_1$, which allows us for simplicity to choose $w^\theta=0$, which implies also $w^{(\theta)}=0$. By expanding
$\dot{w}_\mu=0$ for the other components of the reference vector along the under investigation geodesic trajectory we get:
\begin{eqnarray}
 &&\dfrac{dw_{t}}{d\tau}=\dfrac{M }{r^2\Delta}\Big\{\mathcal{U}^r \Big[(r^2+a^2)w_t+aw_{\phi}\Big]-Mr^2\mathcal{U}_t w^r\Big\},\  \\
 && \dfrac{dw_{r}}{d\tau}=\dfrac{1}{\Delta^2}\Big[r(a^2-Mr)\mathcal{U}^rw^r+\mathcal{U}_t\Big(M(a^2-r^2)+a(M-r)w_{\phi}\Big)\Big] \\
 &&\dfrac{dw_{\phi}}{d\tau}=\dfrac{1}{r^2\Delta}\Big\{\mathcal{U}^r \Big[-aM(a^2+3r^2)w_t+(r^3-2Mr^2-a^2M)w_{\phi}\Big]+ar^2(r+M)\mathcal{U}_tw^r\Big\},
 \end{eqnarray}
A simple combination between the $t$ and $\phi$ component reveals that: 
\begin{equation}
a(r+M) \dfrac{dw_{t}}{d\tau}+\dfrac{dw_{\phi}}{d\tau}=\dfrac{M}{r}\mathcal{U}^{r}\Big(w_{\phi}+a w_{t}\Big).    
\end{equation}
It is possible to rewrite the above relation into a more convenient form as:
\begin{equation}\label{eq:wrtCon}
    \frac{M}{r}(w_{\phi}+aw_t(1+r/M))=C_{rl}=\text{constant}.
\end{equation}
Note that in the tetrad field formulation, we have
\begin{align}\label{eq:wphitetr}
   w^{(\phi)}=w_{(\phi)}=\dfrac{1}{r}(w_{\phi}+a w_t).
\end{align}
We will discuss the implications of the above two relations below. For this discussion we need also the relation%
\begin{equation}\label{eq:wrtNorm}
 [w^r]^2=\dfrac{1}{r^2M^2}\Big\{\Big[(M^2-a^2)w^2_t+2aC_{rl}w_t-(M^2+C^2_{rl})\Big]r^2+2Mr [C^2_{rl}+M^2-aC_{rl}w_t]-a^2M^2\Big\},
\end{equation}
which comes from the fact that $w^\mu w_\mu=-1$ and \ref{eq:wrtCon}.

Our first aim is to check whether it is possible to have a vanishing hidden momentum in this setup. We start with the \OKS~\SSC, which introduces the following relations 
\begin{eqnarray}
S^{(t)(r)}=-S^{(t)(\phi)}w^{(\phi)}/w^{(r)}, S^{(t)(\theta)}=(S^{(r)(\theta)}w^{(r)}-S^{(\theta)(\phi)}w^{(\phi)})/w^{(t)}, S^{(t)(\phi)}=S^{(r)(\phi)}w^{(r)}/w^{(t)}.\label{eq:SPIN_SSC}
\end{eqnarray}
Using the above expressions, we expand $K^{(\alpha)(\beta)}w_{(\beta)}$, and arrive at
\begin{eqnarray}
K^{(t)(\beta)}w_{(\beta)}&=&\dfrac{3MC_{S^2}w^{(\phi)}}{r^3 w^{(t)}} \Big(S^{(r)(\theta)}S^{(\theta)(\phi)}w^{(r)}+(S^{(r)(\phi)}-S^{(\theta)(\phi)})w^{(\phi)}\Big), \nonumber \\ K^{(r)(\beta)}w_{(\beta)}&=& \dfrac{3MC_{S^2}w^{(\phi)}}{r^3[w^{(t)}]^2}\Big(S^{(r)(\theta)}S^{(\theta)(\phi)}[w^{(t)}]^2+S^{(r)(\theta)} w^{(r)}w^{(\phi)}\Big), \nonumber \\
K^{(\theta)(\beta)}w_{(\beta)}&=& \dfrac{3MC_{S^2}S^{(r)(\phi)}w^{(\phi)}}{r^3 [w^{(t)}]^2}\Big(S^{(r)(\theta)}([w^{(t)}]^2-[w^{(r)}]^2)+S^{(\theta)(\phi)}w^{(r)}w^{(\phi)}\Big), \nonumber \\
K^{(\phi)(\beta)}w_{(\beta)}&=& -\dfrac{3C_{S^2}Mw^{(\phi)}}{r^2[w^{(t)}]^2}\Big([S^{(\theta)(\phi)}]^2 [w^{(t)}]^2+[S^{(r)(\phi)}]^2 (-[w^{(t)}]^2+[w^{(r)}]^2)\Big).
\end{eqnarray}
The above relations identically become zero if we chose $w^{(\phi)}=0$. For $w^{(\phi)}=0$, \ref{eq:wrtCon} and \ref{eq:wphitetr} lead to $a~w_t=C_{rl}$ and $w_\phi=-C_{rl}$. Note, that for $a=0$ we get back to the \S~case, for which we have chosen $w_\phi=0$. Moreover, for the above choices of  $w_t$ and $w_{\phi}$, \ref{eq:wrtNorm} reduces to
\begin{equation}
[w^r]^2=\dfrac{C^2_{rl}}{a^2}-\dfrac{\Delta}{r^2}.    
\end{equation}
When we compare $w^r$ with $\mathcal{U}^r$ from \ref{eq:4-velocity_03}, we see that by choosing $C_{rl}=-a \tilde{E}_g=-\tilde{L}_{\rm z}$, they become the same. This implies that $w^\mu = \mathcal{U}^\mu$, i.e. the \OKS~\SSC~and the \P~\SSC~centroids coincide. Since we have a vanishing hidden momentum also the \TD~\SSC~coincides with the other two.

We stress that $C_{rl}=-a \tilde{E}_g$ is just an initial condition choice for the $w^\mu$. Namely, if we assume that initially $w^{\mu}$ is equal to $\mathcal{U}^{\mu}$, and after that it follows its own evolution equation, i.e., $\dot{w}^{\mu}=0$, we found that both $ d w_{t}/d\tau$ and $ d w_{\phi}/d\tau$ identically vanish. This would essentially mean that $w_t$ and $w_{\phi}$ remain constant throughout the evolution. {\it Actually, if $w^{\alpha}$ is initially set to be equal to $\mathcal{U}^{\alpha}$, it will remain equal to $\mathcal{U}^{\alpha}$ for the entire evolution}. We confirmed this also numerically. Hence, setting $w^{(\phi)}=0$ and $C_{rl}=-a \tilde{E}_g$ is equivalent to setting initially $w^\mu=\mathcal{U}^\mu$. We can now state that $w^{(\phi)}=w^{(\theta)}=0$ is a sufficient condition to have a vanishing hidden momentum in the present setup, as it is in \ref{sec:PQS} for the \S~case.

With the issue of choosing the vector $w^{\mu}$ settled, we now get back to the \MPD~equations. Having a vanishing hidden momentum allows us to use again \ref{eq:MPD_PG}, but now for the Kerr case. By using the tetrad field to express \ref{eq:MPD_PG}, we analyse the equations component after component. 
\paragraph{Time component:}
For the $(t)$ component  one immediately obtains $R^{(t)}_{(\mu)(\nu)(\gamma)}\mathcal{U}^{\mu}S^{(\nu)(\gamma)}=0$, and we get $\dot{m}\mathcal{U}^{(t)}=-(1/6)J^{(\alpha)(\beta)(\gamma)(\delta)}\nabla^{\mu}R_{(\alpha)(\beta)(\gamma)(\delta)}$. The expressions for the quadrupole term and $\dot{m}$ are given below:
\begin{eqnarray}\label{eq:mdot_Kerr}
&&\dfrac{1}{6}J^{(\alpha)(\beta)(\gamma)(\delta)}\nabla^{(t)}R_{(\alpha)(\beta)(\gamma)(\delta)}=-\dfrac{3aMC_{S^2}}{r^5}S^{(r)(\theta)}S^{(\theta)(\phi)}, \nonumber \\
&&\dot{m}=\dfrac{3C_{S^2}M\sqrt{\Delta}}{2r^5[w^{(t)}]^2}\Big\{\mathcal{U}^{(r)}\Big([(S^{(r)(\theta)})^2+(S^{(r)(\phi)})^2]-2 [S^{(\theta)(\phi)}]^2 [w^{(t)}]^2\Big)\Big\}.
\end{eqnarray}
It is easy to notice that apart from the tetrad notation, the expression for $\dot{m}$ remains identical to the \S~case, as shown in \ref{eq:mdot}. 

\paragraph{Polar component:}
For the $\theta$-component, we have $\mathcal{U}^{(\theta)}=0$, and it simply reduces to 
\begin{equation}
\dfrac{3MC_{S^2}\sqrt{\Delta}S^{(r)(\phi)}S^{(\theta)(\phi)}}{r^5}=0, 
\end{equation}
and we end up with $S^{(r)(\phi)}=0$ or $S^{(\theta)(\phi)}=0$.
%%%
\paragraph{Azimuthal component:}
%%%
For the $\phi$-component, we obtain
%%%
\begin{equation}
\dfrac{3MC_{S^2}\sqrt{\Delta}S^{(r)(\theta)}S^{(\theta)(\phi)}}{r^5}=0.   
\end{equation}
%%%
and here, we either have $S^{(r)(\theta)}=0$ or $S^{(\theta)(\phi)}=0$. Note that to satisfy both the $\theta$ and $\phi$ components, we must have $S^{(\theta)(\phi)}=0$, and $S^{(r)(\theta)}$ and/or $S^{(r)(\phi)}=0$. However, if we substitute $S^{(\theta)(\phi)}=0$ into the first expression of \ref{eq:mdot_Kerr}, it vanishes. With this, we arrive at $\dot{m}=0$, which translates into $S^{(r)(\theta)}=S^{(r)(\phi)}=0$. So, all the components of the spin tensor identically vanish. \textit{This concludes that the geodesic solution found in \ref{sec:PDK} cannot hold when the body has a spin-induced quadrupole moment.}

%\textit{This concludes that  an extended object with dipole and spin-induced quadrupole, can not follow a geodesic trajectory with $L_{\rm z}=a~E_{\rm g}$ in a Kerr spacetime, if $w^{\mu}$ is set equal to $\mathcal{U}^{\mu}$ initially.}

\section{Conclusions} \label{sec:remarks}

In this work, we took advantage of the vanishing hidden momentum property of the \OKS~\SSC~\cite{Kyrian:2007zz,Costa:2014nta}, which allows the four-momentum to be parallel to the four-velocity, to find common solutions between geodesic and \MPD~equations in \S~and Kerr. Namely, we showed that in the \S~spacetime, the only possible solution for a pole-dipole body to follow a geodesic trajectory is the radial motion. It should be stressed, that this solution holds not only for the \OKS~condition, but for the \P~and \TD~\SSC s as well, which implies that for the radial trajectories the three centroids coincide. However, this does not imply that either the \TD~or the \P~condition can not have any other geodesic solution apart from the radial one. This is yet to be confirmed.

On the equatorial plane of Kerr spacetime, we established that for a pole-dipole body under \OKS~\SSC, there is a unique solution allowing geodesic trajectories to exist. For these trajectories there is a special relation between the geodesic orbital angular momentum $L_{\rm z}$ and the geodesic energy $E_{\rm g}$, which reads $L_{\rm z}=a E_{\rm g}$. It has been shown \cite{chandrasekhar1998mathematical} that these geodesic trajectories in Kerr have similar features with the radial geodesics ($L_{\rm z}=0$) found in the \S~background. The solution we found is also valid for the \P~and \TD~\SSC s, but the uniqueness has not been proven for these \SSC s in our work.

In the case of a pole-dipole-quadrupole body, we discussed the condition $K^{\mu\nu}w_\nu=0$ under which the hidden momentum can vanish for the \OKS~\SSC~as in the pole-dipole case. For the specific cases examined in this work, we employed solely spin-induced quadrupole moment. By considering a particular choice of the reference vector $w^\mu$, we confirmed that  $K^{\mu\nu}w_\nu=0$ is valid on a radial trajectory in the \S~spacetime and on an equatorial orbit in Kerr for which $L_{\rm z}=a E_{\rm g}$ holds. But, we speculate that in general, for a pole-dipole-(spin induced) quadrupole body under this supplementary condition, the hidden momentum may not vanish. Finally, we demonstrated that in the above two examples the extended body under \OKS~\SSC~cannot follow a geodesic trajectory.

\section*{Acknowledgement}

S.M. wish to acknowledge Narayan Banerjee for some useful interactive sessions at the earlier stage of this work. The authors would like to express their gratitude to L. Filipe O. Costa for the clarification he has provided and for pointing out some important consequences of the present work. They would also like to thank the Inter-University Centre for Astronomy and Astrophysics (IUCAA), Pune, India, where parts of this work were carried out during academic visits. The authors wish to thank the referees for their constructive criticism on the article. S.M. and G.L.-G. have been supported by the fellowship Lumina Quaeruntur No. LQ100032102 of the Czech Academy of Sciences.

\section*{Data availability}
Data sharing not applicable to this article as no datasets were generated or analysed during the current study.

\bibliographystyle{utphys1}
\bibliography{References}

\end{document}